\documentclass[prd,aps,showpacs,12pt,amssymb,preprintnumbers,amsmath,amsfonts]{revtex4-1}
\usepackage{graphicx}

\newcommand{\be}{\begin{equation}}
\newcommand{\ee}{\end{equation}}
\newcommand{\ba}{\begin{eqnarray}}
\newcommand{\ea}{\end{eqnarray}}
\newcommand{\nn}{\nonumber}
\newcommand{\beq}{\begin{equation}}
\newcommand{\eeq}{\end{equation}}

\newcommand{\Eq}[1]{Eq.~\eqref{#1}}

\newcommand{\hq}{{\hat q}}

\newcommand{\hel}{{\lambda}}
\newcommand{\Tr}{{\rm{Tr}}}
\newcommand{\p}{{\bf{p}}}
\newcommand{\q}{{\bf{q}}}
\newcommand{\vv}{{\bf{v}}}
\newcommand{\kk}{{\bf{k}}}

\usepackage{slashed}

\usepackage{hyperref}

\date{\today}

\begin{document}

\title{Measuring chiral imbalance with collisional energy loss} 
\author{Stefano Carignano}
\email{carignano@ice.cat}
\affiliation{Instituto de Ciencias del Espacio (ICE, CSIC) \\
C. Can Magrans s.n., 08193 Cerdanyola del Vall\`es, Catalonia, Spain
and \\
 Institut d'Estudis Espacials de Catalunya (IEEC) \\
 C. Gran Capit\`a 2-4, Ed. Nexus, 08034 Barcelona, Spain
}
\author{Cristina Manuel}
\email{cmanuel@ice.csic.es}
\affiliation{Instituto de Ciencias del Espacio (ICE, CSIC) \\
C. Can Magrans s.n., 08193 Cerdanyola del Vall\`es, Catalonia, Spain
and \\
 Institut d'Estudis Espacials de Catalunya (IEEC) \\
 C. Gran Capit\`a 2-4, Ed. Nexus, 08034 Barcelona, Spain
}

\begin{abstract}
{We compute the collisional energy loss of an {energetic} massive fermion crossing a chiral plasma at finite temperature
characterized by an imbalance between the populations of left-handed and right-handed fermions. We find a new contribution to the 
energy loss which is proportional to the helicity of the test fermion and depends on the amount of 
chiral imbalance in the plasma.  We then compute the difference between the energy loss of a fermion with the two  opposite helicities,
to assess whether this could be used to quantify the chiral imbalance in the plasma. We find that the leading contribution to these helicity-dependent energy loss contributions comes from the exchange of hard photons (or gluons for QCD) with the medium constituents, and in some scenarios can become comparable to the leading-order result for a plasma without any chiral imbalance. We also evaluate the contribution arising from soft photon exchange, which is a subleading effect, and requires regularization.
We illustrate how dimensional regularization is a well suited prescription to be applied to these energy loss computations.
}
\end{abstract}

\maketitle

\section{Introduction}

The measurement of the energy loss for a jet propagating through a high-temperature plasma
is one of the most prominent quantities which can be used to characterize the properties of matter 
in scenarios such as heavy-ion collision experiments (see eg. \cite{dEnterria:2009xfs,CasalderreySolana:2007zz,Majumder:2010qh,Qin:2015srf} for recent reviews).
Of particular interest in such a context is the energy loss of a heavy parton produced in the early stages of the collision, which crosses the quark-gluon plasma
interacting with the medium constituents.

Recently, it has been proposed that the quark-gluon plasma created in a heavy-ion collision could exhibit an imbalance between populations of left-handed and right-handed fermions, 
 giving rise to
several new macroscopic
phenomena (see \cite{Kharzeev:2013ffa,Kharzeev:2015znc,Huang:2015oca} for reviews).

Aside from such anomalous transport phenomena,  the presence of a chiral imbalanced system should affect the interaction of an energetic fermion with the medium, and in particular its collisional energy loss due to interaction with the constituents of the plasma. This feature has already appeared in a calculation of the damping rate of a massless fermion in an imbalanced dense system at zero temperature  \cite{Carignano:2018thu}, where it was found that the interaction with the medium particles mediated by soft photons distinguishes between different photon circular polarizations and depends on the chirality of the test particle. 
This in turn suggests us that a closer investigation of the energy loss might even help shed some light on the amount of chiral imbalance originated in a heavy-ion collision experiment.

In order to investigate this aspect in more detail, 
in this work we compute the collisional energy loss of an {energetic} massive fermion traversing a chiral plasma at finite temperature $T$ made of
unequal populations of left-handed and right-handed massless fermions. {We consider the fermion energy $E$ to be much larger than the temperature, $E \gg T$.}  The imbalance is parametrized by the chiral chemical potential $\mu_5 = \mu_R-\mu_L$, where $\mu_{R/L}$ refer to the chemical potential associated with right/left handed fermions of the medium,
respectively. {We start by considering 
 an electromagnetic plasma, and later on we discuss how to  generalize our main results to QCD.}

%In the presence of a chiral chemical potential both parity $P$ and $CP$, where $C$ is charge conjugation, are broken. This should then be reflected
%in the energy loss, which will turn out to depend on the helicity of the heavy fermion.
%We shall focus our efforts in computing this helicity contribution, which is subleading in a $1/E$ expansion, where $E$ is the energy of the fermion. 

It is worth recalling that 
 helicity is not a Lorentz invariant quantity. Our computation is carried out in the frame at rest with the plasma, but it should be possible to generalize it to a more convenient lab frame.

The calculation of the collisional energy loss involves considering contributions for both hard and soft momenta of the exchanged photon (recall that hard and soft refer to scales of order $T$ and $eT$, respectively, where $e$ is the gauge coupling constant, and that the soft scales 
 require the resummation of hard thermal loops \cite{Braaten:1991jj,Peigne:2007sd}). For the leading contribution in a plasma without chiral imbalance, both contributions when taken separately exhibit divergencies which cancel when adding them up. 
This separation of scales is typically performed using a sharp momentum cutoff, a procedure which in principle spoils gauge invariance. In this work, in order to calculate the 
new contributions arising in a chiral imbalanced plasma we will follow the same philosophy, but employing dimensional regularization (DR) to regularize all intermediate results. As an illustrative example, in Appendix A we repeat the known calculation of the leading contribution to the energy loss for a plasma without chiral imbalance employing DR. We stress that DR is a perfectly suited regularization method in the presence of power-like divergences, as those we find in our computations. 

We find that the leading contribution to the helicity-dependent energy loss arises from the exchange of hard photons with the medium constituents, and
it does not exhibit any sort of divergence. We also find a contribution to the helicity-dependent energy loss as arising from the exchange of soft photons, which however turns out to be
perturbatively suppressed as compared to the hard contribution.

{Let us finally mention that in a recent publication the collisional energy loss in a chiral medium has been considered, by studying the electromagnetic fields created by a moving electrical charge  in the presence of the chiral magnetic current \cite{Hansen:2020irw,Tuchin:2018sqe}. The effective field  equations  considered in that manuscript however neglect thermal effects, and only  describe the physics of chiral plasmas in the static limit,  as  can been shown by evaluating the polarization tensors of the chiral plasma used in our manuscript.}

This paper is structured as follows: in Sec. \ref{sec:hard}  we compute the {leading} hard contribution to the helicity-dependent collision rate, while in Sec. \ref{sec:soft} we compute the {leading} soft one. We discuss
our results in Sec. \ref{sec:conclusions}.  {In  App.~\ref{app:DR}  we repeat the known calculation of the collision rate in a regular QED plasma without chiral imbalance, employing DR, the soft sector is considered in App.~\ref{app-soft}, while 
App.~ \ref{app-hard} considers the hard sector.  We provide in App.~\ref{app:integralsDR} some of the DR integrals that we used in our computations. }

We work with natural units $\hbar = c = 1$ and metric $g^{\mu\nu} = { diag}(1,-1,-1,-1)$. We denote four-momenta with capital letters, $K^\mu = (k^0, {\bf k})$, and the modulus of three-momenta as $k=\vert {\bf k}|$.

\section{Hard contribution to the collisional energy loss of a massive fermion in the chiral plasma}
\label{sec:hard}

\begin{figure}[h]
\begin{center}
\includegraphics[height=2.5cm]{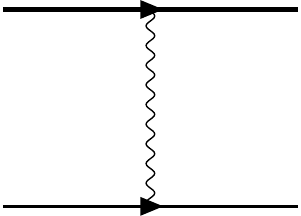} 
\includegraphics[height=2.5cm]{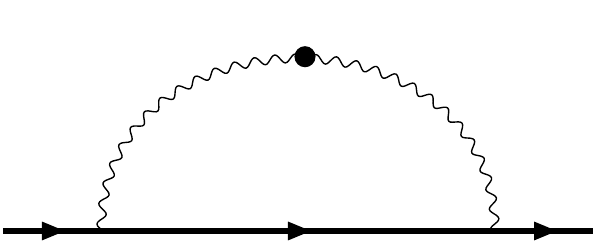} 
\caption{Feynman diagrams for the energy loss calculation. Left: scattering of a heavy fermion (thick line) with a medium fermion via a photon exchange.  Right: heavy fermion self-energy. The dot denotes a resummed medium photon. } 
\label{fig:diagrams}
\end{center}
\end{figure}

Let us start by computing the damping rate of a massive particle with mass $M$, momentum $\p = \vv E $, energy  $E = \sqrt{p^2 + M^2}$ and a given helicity $\hel$, 
due to collision with massless fermions in the chiral plasma, {characterized by a temperature $T$ and chiral chemical potential $\mu_5$}. The relevant scattering diagram is given by
Fig. \ref{fig:diagrams} left, depicting the scattering of a heavy fermion of a given helicity $\hel$ with a massless one in the medium
with chirality $\chi$ via photon exchange. 

The corresponding matrix element squared is given by
\begin{align}
|{\cal M}|^2_{(\lambda\chi)} & = e^4 D_{\mu\nu}(Q) D_{\alpha\beta}^\dagger(Q) {\rm Tr} \left[ {\cal P}_{(\hel,{\bf p})} (\slashed{P} +M) \gamma^\mu (\slashed{P'} +M) \gamma^\alpha \right]
{\rm Tr} [P_{(\chi)}\slashed{K'}\gamma^\nu \slashed{K} \gamma^\beta]  \,,
\end{align}
where $D_{\mu\nu}$ is the photon propagator, 
written in terms of energy and momentum transfer of the
collision $Q^\mu=(\omega, {\bf q})$, which is assumed to be hard, 
\be
{\cal P}_{(\hel,{\bf p})} = \frac{1}{2}\left( 1 + \hel \gamma^5 \gamma^0 \vec\gamma \cdot \hat {\bf p} \right) \ , \qquad  \hel = \pm
\label{eq:helProj}
\ee
is the helicity projector and 
\be
P_{(\chi)} = \frac{1 + \chi \gamma_5}{2} \ , \qquad \chi = \pm
\ee
is the chirality projector. Note that for massless fermions, helicity and chirality agree, but this is not so otherwise.

The first trace, which depends on the helicity of the massive fermion, is given by
\ba
\Tr  \left[{\cal P}_{(\hel,{\bf p})}  (\slashed{P} +M) \gamma^\mu (\slashed{K} +M) \gamma^\nu \right]  = 
2 \left \lbrace \left[ P^\mu K^\nu + P^\nu K^\mu  + (M^2 - P\cdot K) g^{\mu\nu} \right]  \right. \nonumber\\
 \left. - i \hel {\hat p}^i \left[  K_\beta \left(p^i \epsilon^{0\mu\beta\nu} - p^0 \epsilon^{i\mu\beta\nu}  \right) + M^2 \epsilon^{0i\mu\nu} \right] \right\rbrace \,,
 \label{eq:traceHel}
\ea
 while the second one, related to the 
chiral fermion in the medium, is given by 
\be
{\rm Tr} [P_{(\chi)} \slashed{K'}\gamma^\nu \slashed{K} \gamma^\beta] = 
2 \left\{ \left[ K'^\nu K^\beta+ K'^\beta K^\nu  -(K'\cdot K) g^{\nu\beta} \right] + i \chi K_\eta K'_\rho \epsilon^{\nu\beta\eta\rho} \right\} \, .
\label{eq:traceChi}
 \ee

 Splitting the traces into symmetric and antisymmetric pieces, their product will give two contributions:  
  \begin{align}
|{\cal M}|^2_{(\lambda\chi)}  = &  4 e^4 D_{\mu\nu}(Q) D_{\alpha\beta}^\dagger(Q) \times \nn\\
&   \Big\{\left[ P'^\mu P^\alpha+ P'^\alpha P^\mu  + (M^2 - P'\cdot P) g^{\mu\alpha} \right]
  \left[ K'^\nu K^\beta+ K'^\beta K^\nu  - (K'\cdot K) g^{\nu\beta} \right] \nn\\
  & -\lambda\chi \, K_\eta K'_\rho \epsilon^{\nu\beta\eta\rho} \hat{p}^i \left[P'_\sigma p^i \epsilon^{0\mu\sigma\alpha} - P'_\sigma p^0 \epsilon^{i\mu\sigma\alpha} + M^2 \epsilon^{0 i \mu\alpha} \right]    \Big\} \,.
\end{align}
The first contribution gives the usual result known from the literature
\cite{Braaten:1991jj,Peigne:2007sd}, whereas the second {one} depends both on the helicity of the heavy fermion as well as the chirality of the medium fermion. 
In a plasma where there is the same population of
left and right handed fermions this latter contribution vanishes 
  in the final computation of the damping rate or
energy loss when summing over chiralities, but this is not the case otherwise. 

The damping rate is given by integrating the matrix element squared over phase space  and summing over the  chiralities of the fermions in the plasma.
We shall concentrate in the following in the interaction rate, which is related to the damping as $\Gamma_\hel = 2 \gamma_\hel$.
At leading order in the QED coupling constant we get
 \begin{align}
\Gamma^{\rm hard}_{\hel} &=   \frac{1}{E} \int \frac{d^3 p'}{(2\pi)^3} \frac{1}{{2}E'} \int \frac{d^3 k}{(2\pi)^3}  \sum_{\chi=\pm} \frac{n_\chi(k)}{2k}
\int \frac{d^3 k'}{(2\pi)^3} \frac{1 - n_\chi(k' )}{2 k'}
 (2 \pi)^4 \delta^4( P+K -P' -K')
   | {\cal M}|_{(\hel\chi)}^2  \nn\\
 & = \frac{8 e^4}{E} \int \frac{d^3 p'}{(2\pi)^3} \frac{1}{{2}E'} \int \frac{d^3 k}{(2\pi)^3}  \sum_{\chi=\pm} \frac{n_\chi(k)}{2k}
\int \frac{d^3 k'}{(2\pi)^3} \frac{1 - n_\chi(k' )}{2 k'}
 (2 \pi)^4 \delta^4( P+K -P' -K') \nn\\
& \times \frac{E^2}{Q^4} \Bigg\{ \Big[2(k -\vv\cdot\kk)(k'-\vv'\cdot\kk') + \frac{Q^4}{4E^2} + \frac{M^2Q^2}{2E^2} \Big]
 +  \lambda\chi
 \Big(- \frac{Q^2}{2E}\Big) \Big[v(k+k') - \hat{\vv}\cdot(\kk+\kk') \Big] \Bigg\} \,,
\label{eq:dampingHard}
\end{align}

where
\be
 n_\chi(k) = \frac{1}{\exp((k- \chi\mu_5)/T) + 1} \ , \qquad \chi = \pm   \ , 
\ee
 is the occupation number for a fermion of  chirality $\chi$. Thus, we assume that in the plasma there is only a chiral chemical potential and no baryonic chemical potential.

 In the last row of \Eq{eq:dampingHard}
 one can recognize the result for a symmetric plasma \cite{Braaten:1991jj,Peigne:2007sd}, 
 followed by the first non-vanishing correction in a chiral plasma, which is proportional to the product $\hel\chi$ of the helicity and the chirality of the heavy and the light medium fermion, {respectively}.
  Note that this correction is $1/E$ suppressed compared to the leading result.

Let us now consider a very energetic fermion ($E \gg T$), and move to the energy loss $-dE/dx$,
which can be obtained by multiplying the integrand of the damping rate by a factor $(E-E')/v$ \cite{Braaten:1991jj}.

 If one simply takes the hard contribution computed in this section, the energy loss at leading order turns out to be infrared divergent.  
This is cured by appropriately taking into account the contribution from soft momenta in the exchanged photon, which requires a proper resummation. 
In \cite{Braaten:1991jj,Peigne:2007sd}, this is implemented by introducing
a cutoff which separates the hard and soft contributions of the computation.
Even though the final result does not depend on the cutoff parameter, this regularization method has some clear drawbacks, including an ambiguity in the 
choice of the cutoff itself (as a matter of fact, two different ways have been employed in \cite{Braaten:1991jj} and \cite{Peigne:2007sd}). 
An alternative approach
would be to employ 
 dimensional regularization, which has several advantages over the cutoff regularization, as it preserves the gauge invariance. 
We illustrate the method in Appendix \ref{app:DR}.

On the other hand, the new contribution we compute in the following, which is proportional to the helicity of the external fermion, turns out to be finite and does not require any regularization. 
In order to focus on the new chirality and helicity-dependent effects we are computing, let us consider the difference 
 between the energy loss associated with the two opposite helicities:
\begin{align}
\Delta^{\rm hard} &\equiv  \left(- \frac{ d E_{\lambda = +}}{d x}\right)  \Bigg |_{\rm hard} -  \left(- \frac{ d E_{\lambda = -}}{d x} \right)
\Bigg |_{\rm hard} \,.
\end{align}

\begin{align}
 \Delta^{\rm hard} & = \frac{16 e^4}{vE}
  \int \frac{d^3 p'}{(2\pi)^3} \frac{1}{{2}E'} \int \frac{d^3 k}{(2\pi)^3}  \sum_{\chi=\pm} \chi \frac{n_\chi(k)}{2k}
\int \frac{d^3 k'}{(2\pi)^3} \frac{1 - n_\chi(k' )}{2 k'}  
 (2 \pi)^4 \delta^4( P+K -P' -K') \nn\\
& \times (E-E') \frac{E^2}{Q^4} \Big\{ 
 \Big(- \frac{Q^2}{2E}\Big) \Big[v(k+k') - \hat{\vv}\cdot(\kk+\kk') \Big] \Big\} \,.
\label{eq:helCorrHard}
\end{align}
At this point one can proceed and eliminate the $p'$ integral with the $d=3$ spatial delta functions.
 The remaining delta of energy conservation
can be approximated, again to leading order in a $1/E$ expansion, as
$\delta ( \omega - {\bf v} \cdot \q - Q^2/(2E))$, and neglecting higher order terms we can 
drop the $n_F(k')$ above due to the symmetries of the integrand (the expression in the second row of \Eq{eq:helCorrHard} is antisymmetric under the exchange $\kk\leftrightarrow \kk'$).

By further introducing
\be
1 = \int d^d q\,\delta^{(d)}( \q +{\bf k} - {\bf k'}) \int d \omega \, \delta( \omega + k -k') \,,
\ee
we can eliminate the integral in $k'$ and write it in terms of $q$, the momentum transfer.

At this point, one can use the delta function to carry out one of the angular integrals, and perform also all the trivial angular integrations.
More specifically, we write $\delta ( \omega + k -|{\bf k}+ \q|) = 2 |{\bf k}+ \q| \delta (Q^2 + 2 k \omega - 2 \kk\cdot \q ) $ and use it to replace the 
$ \kk\cdot \q$ pieces in our expression.
{Performing an angle average over the directions of the incoming fermion and
 keeping only the leading $1/E$ contribution we arrive at 
\begin{align}
\Delta^{\rm hard}&  = - \frac{e^4}{4\pi^3v^3 E}
 \int_0^\infty dk [n_+(k) - n_-(k)] \nn\\
&\times  \int_0^\infty dq \int_{\omega_-}^{\omega_+} d\omega \frac{\omega}{Q^2} 
 \Big[ \Big(v^2 - \frac{\omega^2}{q^2}\Big)(2k+ \omega) \Big]
  \Theta(|q - k| \leq |\omega + k| \leq q + k) \,, 
\end{align}
where $\Theta$ is the step function, and
with $\omega_{\pm} = E - \sqrt{E^2 + q^2  \mp 2Evq}$.  As in \cite{Peigne:2007sd}, we can again split the integrals in two regions, 
\be
\Delta^{\rm hard}
= - \frac{e^4}{4\pi^3v^3 E} \Big[ {\cal I}_1 + {\cal I}_2 \Big]\,.
\label{eq:dhard}
\ee

The first integral is given by 
\begin{align}
{\cal I}_1 &= 
 \int_0^\infty dk [n_+(k) - n_-(k)] \int_0^{\frac{2k}{1+v}} dq \int_{-vq}^{vq} d\omega \frac{\omega}{Q^2} 
  \Big[ \Big(v^2 - \frac{\omega^2}{q^2}\Big)(2k+ \omega) \Big] 
\nn\\
 & = {-} \frac{2}{(1+v)^2}\frac{2}{3}\Big[ v(2v^3-3)-\frac{3}{2}(v^2-1)\log\Big(\frac{1+v}{1-v}\Big)\Big] \frac{\mu_5}{3} \Big(\mu_5^2 + \pi^2 T^2 \Big) \,.
\end{align}

The second contribution, which amounts to
\be
{\cal I}_2 = 
 \int_0^\infty dk [n_+(k) - n_-(k)] \int_{\frac{2k}{1+v}}^{q_{\rm{\max}}} dq \int_{q-2k}^{\omega_+} d\omega \frac{\omega}{Q^2} 
  \Big[ \Big(v^2 - \frac{\omega^2}{q^2}\Big)(2k+ \omega) \Big]  \,,
  \label{eq:i2}
\ee
with $q_{\rm{\max}} = 2k(E+k)/(E(1-v)+2k)$, needs some more careful treatment.

 Following again \cite{Braaten:1991jj,Peigne:2007sd}, we can treat separately the regime $E \ll M^2/T$,
 where we can approximate $\omega_+ \simeq v q$, $q_{\rm max} \simeq 2k/(1-v)$, obtaining 
\begin{align}
{\cal I}_2^{(E \ll M^2/T)} 
& = {-}  \frac{1}{6 (1 + v)^2} \Big[\frac{2 v}{v-1} (-21 + 3 v + 23 v^2 + 7 v^3)  \nn\\
 & +   3 (v-1) (7 + 13 v + 9 v^2 + 3 v^3) \log\frac{1 + v}{1 - v}\Big]
  \frac{\mu_5}{3} \Big(\mu_5^2 + \pi^2 T^2\Big) \,,
  \label{eq:i2EllM}
\end{align}
whereas for $E \gg M^2/T$, corresponding to the $v\to 1$ case, the appropriate limit is $\omega_+ \simeq q, q_{\rm max} \simeq E$, leading to
a much more involved expression.

We can however obtain an analytical expression for the $v\to 1$ limit
\begin{align}
{\cal I}_2^{(v = 1)}  & = \int_0^\infty dk  [n_+(k) - n_-(k)]  \frac{2}{3} k \Big(3 E - 5 k + 2 \frac{k^2}{E} \Big) \nn\\
& =  2 E T^2 \Big[ Li_2(-e^{-\mu_5/T})- Li_2(-e^{\mu_5/T}) \Big] + \dots 
\end{align}
where the dots denote subleading terms in a $1/E$ expansion, and $Li_2$ denotes the dilogarithm function.  We note that for large $E$ this is a leading correction.
If we further expand in the limit $\mu_5 \ll T$, we obtain
\be
{\cal I}_2^{(v=1)} \simeq 2 E T \mu_5 \log 2 + {\cal O}\Big(\frac{\mu_5}{T}\Big)  \,.
\ee

Putting together the two contributions in \Eq{eq:dhard}, we reach to the final hard contribution result, valid for $E \ll M^2/T$ 
\be
\label{main-res}
\Delta_{(E \ll M^2/T)}^{\rm hard}  =  \frac{e^4 T^2}{12 \pi} \frac{\mu_5}{E} \Big( 1 + \frac{\mu_5^2}{\pi^2 T^2} \Big)
\Big[ \frac{3 v - 5 v^3}{v^3(v^2-1)} - \frac{3}{2v^3}  (v^2 -1) \log\frac{1+v}{1-v} \Big] \,, 
\ee
while for $v=1$  ( $E \gg M^2/T$ ), in the limit  $\mu_5 \ll T$ the leading contribution is given by 
\be
\label{main-res-v1}
\Delta_{{(v = 1)}}^{\rm hard}  \approx \frac{e^4}{2\pi^3}\mu_5 \, T \, {\log 2} +  {\cal O}\Big(\frac{\mu_5}{T}, \frac{\mu_5}{E} \Big)  \,,
\ee
 which becomes almost comparable with the leading-order result for the collisional energy loss, which is ${\cal O}(e^4 \log(e) T^2)$.
It is also worth recalling that for $v=1$ the helicity of the fermion coincides with the chirality, and it is then a Lorentz invariant.

The other process which could contribute to the collisional energy loss is Compton scattering with medium photons.
One can nevertheless see that, at this order of the calculation, all the relevant contributions which depend on the fermion helicity and the chiral imbalance of the system vanish,
so that this process will not contribute to $\Delta$ {at  the same order as Eqs.(\ref{main-res}) and (\ref{main-res-v1}).}

\section{Soft contribution to the collisional energy loss of a massive fermion in the chiral plasma}
\label{sec:soft}

Let us now focus on the contribution to the energy loss coming from soft momentum photons,
mainly focusing on contributions that depend on the helicity of the test fermion.
 The most convenient way to calculate
these contributions is to start from the computation of the fermion damping rate \cite{Weldon:1983jn}
\be
\gamma_\hel = -\frac{1}{2E} {\rm Tr} \left[ {\cal P}_{(\hel,{\bf p})} (\slashed{P} +M) {\rm Im} \Sigma(P) \right] \Big\vert_{p_0 = E} \,,
\label{eq:dampinghel}
\ee
where  ${\cal P}_{(\hel,{\bf p})}$ are the helicity projectors defined in \Eq{eq:helProj},
and the fermion self-energy $\Sigma(P)$ (Fig. \ref{fig:diagrams} right) can be written in terms of four independent scalar functions $\Sigma^\hel_s(P)$
\be
\Sigma(P) = \sum_{\lambda= \pm} \sum_{s= \pm} {\cal P}_{(\hel,{\bf p})} \gamma_0 \Lambda_{(s,{\bf p})} \,\Sigma^\hel_s(P) \,,
\ee
where
 \be
\Lambda_{(\pm,{\bf p})} = \frac{E \pm (\gamma_0 \vec\gamma \cdot {\bf p} + \gamma_0 M )}{2 E}
\ee
 are  particle/antiparticle projectors. One can  check that $\left [{\cal P}_{(\hel,{\bf p})}  ,\Lambda_{(\pm,{\bf p})}  \right] =0$.

The fermion self-energy can be computed {\it e.g.} using the imaginary time formalism, and then analytically continued to Minkowski
space time.  For a  plasma without any
chiral imbalance, the self-energy corrections for the two helicities turn out to be the same \cite{Manuel:2000mk}. In the presence
of a non-vanishing $\mu_5$ however, one can expect the damping rate to depend on the helicity of the fermion, only based on the global symmetries of the system,
as parity is broken.

 If we compute the fermion damping rate in the regime where the photon carries soft momentum, the photon in the one-loop diagram has to be resummed.
If we use the hard thermal loop (HTL) resummed photon propagators, there is still no helicity dependence on the fermion damping rate. However, we find an helicity dependence
if we improve the HTL resummation by considering $\mu_5$ corrections to the resummed photon propagators.

As in \cite{Carignano:2018thu},   we consider that in  Coulomb gauge  
(we ignore gauge dependent pieces here, as they do not contribute to the imaginary part of the fermion self-energy)
the photon propagator can be written as  \cite{Nieves:1988qz}
\be
D_{\mu \nu} (Q) = \delta_{\mu 0} \delta_{\nu 0} D_L(Q) + \sum_{h=\pm} {\cal P}^{T,h}_{\mu \nu} D_T^h(Q) \,,
\ee
where
  $h=\pm$ labels the two circular polarised transverse photon states, left and right,  and   
\be
{\cal P}_{\mu \nu}^{T,h} = \frac 12 \Big(\delta^{ij} - \hq^i \hq^j - i h \epsilon^{ijk} \hq^k \Big) \delta_{\mu i} \delta_{\nu j} \ .
\ee
The resummed longitudinal and transverse propagators  read, with the usual prescription  $\omega \rightarrow \omega \pm i \eta$ for retarded and advanced quantities, respectively,
\be
D_L (\omega, q)= \frac{1}{q^2 + \Pi_L} \ , \qquad  D_T^h(\omega,q) = \frac{1}{\omega^2 -q^2 - \Pi_T - h \Pi_P}\,,
\ee
where 
\begin{eqnarray}
\label{pipiL}
\Pi_L (\omega,q) & = & m^2 _{D}  \left(1- \frac{\omega}{2 q} 
 \,{\rm ln\,}{\frac{\omega+ q}{\omega- q}}  \right)
  \ , \\
 \Pi_T (\omega,q) & = & m^2 _{D} \, \frac{\omega^2}{2  q^2} \left[ 1 + \frac12 \left( \frac{q}{\omega} -
\frac{\omega}{ q} \right) \,  {\rm ln\,} {\frac{\omega+
 q}{\omega- q}} 
\, \right] \ ,
 \label{pipiT}
\end{eqnarray}
are the longitudinal/transverse part of the hard thermal/dense loop photon polarization tensor \cite{LeBellac},
and
 $
 m^2_D = e^2  \left( \frac{T^2}{3} + \frac{\mu^2_5}{ \pi^2} \right)
$
is the Debye mass, 
while 
\be
\Pi_P (Q)=  - \frac{e^2 \mu_5}{ \pi^2}  \frac{\omega^2-q^2}{ q } \Big[ 1 - \frac{\omega}{2  q}   {\rm ln\,} {\frac{\omega+ q}{\omega-q}}
  \Big] \,
\ee  
can be viewed as the anomalous hard dense loop contribution \cite{Laine:2005bt,Akamatsu:2013pjd,Manuel:2013zaa}.

Please also note that by improving the HTL resummed photon propagators by adding the new anomalous contribution, means that we 
are including a correction of order $e$ to the standard HTL result,  as it is a correction $\sim e^2 \mu_5 q_{\rm soft} \sim e\, q^2_{\rm soft}$,   where $ q_{\rm soft}$ is a soft momentum, of order $eT$ or $ e \mu_5$.

Let us point out here that by analyzing the poles of the transverse propagators, one finds a  chiral plasma instability \cite{Laine:2005bt,Akamatsu:2013pjd}.
However, the time scales associated with the instability are relatively large (${t_{\rm ins} \sim  T^2/e^4 \mu^3_5}$, for $\mu_5 \sim T$,  $t_{\rm ins} \sim 1/e^4 \mu_5$), and the computation of the energy loss we carry out is valid for shorter 
time scales.

The spectral functions associated with the longitudinal and transverse gauge field modes
are given by
\begin{align}
 \rho_{L} (Q)  =  &2 \,{\rm Im} \,D_{L}(\omega + i\eta, q) \ ,  \\
  \rho_{T}^h (Q) = &  2 \,{\rm Im} \, D_{T}^h(\omega + i\eta, q)  \ , \qquad h= \pm \ ,
\label{spectral}
\end{align}
respectively.

Using these resummed photon propagators, after some standard manipulations, one ends up with 
the following value of the interaction rate
\begin{align}
\label{Inter-int-damp}
\Gamma_\hel =  & \frac{e^2}{ 2E}  \int \frac{d^3q}{(2\pi)^3} \int d\omega [ 1 + n_B(\omega) ] \delta(E-E' - \omega) \frac{1}{E'} \nonumber\\ 
& \times \Big\lbrace \left[ E E' + E^2 - \p\cdot\q  \right]\rho_L(Q) + \left[ E E' - M^2 -(\p\cdot\hat{\q})^2 + \p\cdot\q  \right] \sum_{h=\pm}\rho_T^h(Q)  \nonumber\\
& - \hel \left[ (E E' -M^2 - p^2)(\hat{p}\cdot\hat{q}) + p q \right]  \sum_{h= \pm} h \rho_T^h(Q) 
 \Big\rbrace \,,
\end{align}
where $ E' = \sqrt{({\bf{p -q}})^2+M^2}$, 
$n_B(\omega) = (exp(\omega/T) - 1)^{-1}$ is the bosonic occupation number and we assumed that $M, E' \gg T$, which allows to neglect the fermionic occupation numbers in our expression. 

The above integral can be analyzed in an expansion on $1/E$. Let us define the velocity vector as ${\bf{v}} = \p/E$
and expand
\be
E' = \sqrt{(\p - \q)^2 + M^2 } \simeq E - vq\cos\theta + \frac{q^2(1-v^2\cos^2\theta)}{2E} + {\cal O} \Big(\frac {1}{E^2}\Big) \ ,
\ee

We can write
\begin{align}
\Gamma_\hel & = \frac{e^2}{4 \pi^2 v} \int_0^\infty  dq q \int_{-q v}^{qv} d\omega  [ 1 + n_B(\omega) ]  \Big\lbrace \Big(1-\frac{\omega}{E} \Big) \rho_L(\omega,q)  \nonumber\\
& + \frac12 \sum_{h=\pm} \rho_T^h(\omega,q ) \Big[ \Big(v^2-\frac{\omega^2}{q^2}\Big)\Big(1  - 
  \frac{\hel h q}{v E}\Big)  - \frac{\omega}{E}\Big( 1 -\frac{\omega^2}{q^2} \Big)  \Big]   \Big\rbrace  \nn\\ 
  & - \frac{e^2(1-v^2) }{8 E \pi^2 v} \int^\infty_0 dq  q^3 \rho_L(q v,q)   \,,
  \label{eq:dampingSoft}
\end{align}
which in the limit of a massless fermion $v\to 1$ reduces to the expressions found in \cite{Carignano:2018thu,Carignano:2019zsh,Carignano:2018gqt}, as expected.

 When $\mu_5=0$  our result agrees with that of Refs.\cite{Braaten:1991jj} and \cite{Peigne:2007sd} up to order $1/E$  (after taking into account a different factor of $2\pi$ in the convention for the definition of the photon spectral functions).
 At order $1/E$ we find the new contributions that only appear in the
presence of chiral imbalance (as otherwise $\rho_T^+ = \rho_T^-$ and they would cancel when summing over polarizations), which depend on both the helicity of the fermion and the circular polarization of the photon.

The energy loss can again be obtained from the damping by multiplying by $\omega/v$ \cite{Braaten:1991jj}.  We have
\begin{align}
- \frac{ d E_\hel}{d x} \Bigg |_{\rm soft} &=  e^2  \int \frac{d^3 q}{(2\pi)^3}  \int d\omega [ 1 + n_B(\omega) ] \delta(E-E' - \omega) \frac{1}{2 E E'} \frac{\omega}{v}\\
& \times \Big\lbrace \left[ E E' + E_p^2 - \p\cdot\q  \right]\rho_L(Q) + \left[ E E' - M^2 -(\p\cdot\hat{\q})^2 + \p\cdot\q  \right] \sum_{h=\pm}\rho_T^h(Q)  \nonumber\\
& - \hel \left[ (E E' -M^2 - p^2)(\hat{p}\cdot\hat{q}) + p q \right]  \sum_{h= \pm} h \rho_T^h(Q) 
 \Big\rbrace \,.
\end{align}

 Focusing again on the difference 
 between the energy loss of two particles with opposite helicities,
\begin{align}
\Delta^{\rm soft} &\equiv \left(- \frac{ d E_{\lambda = +}}{d x} \right)\Bigg |_{\rm soft} -\left(- \frac{ d E_{\lambda = -}}{d x} \right)
\Bigg |_{\rm soft} \,,
\end{align}
we get

\begin{align}
\Delta^{\rm soft} 
& =  \frac{ e^2}{4\pi^2 v^2}  \,\int_0^\infty dq \, q^2 \int_{-v}^v dx  (q x)  [ 1 + n_B (q x) ] 
  \frac12 \sum_{h=\pm} \rho_T^h(qx,q )  \Big(v^2- x^2\Big)\Big( -  2 \frac{h q}{v E}\Big)  \,.
  \label{eq:deltaSoft}
\end{align}

 We can plug in the explicit value of the transverse spectral functions. One has \cite{Carignano:2018thu}
\beq
\frac{\rho^h_{T,{\rm cut}}}{2\pi}(\omega, q)  =
 \frac{ M^2_h \, \, \frac{x}{1-x^2} \, \Theta(1-x^2)}{
\left[ 2 q^2 +  \frac{m_D^2}{1-x^2} -  M^2_h Q_1(x) \right]^2 + \frac{M_h^4 \pi^2 x^2}{4}} \ ,\, \qquad x = \omega/q \,,
\eeq
with 
$M^2_h = m^2_D  -\frac{e^2 \mu_5}{2\pi^2}  h q$, 
\beq
Q_1(x) = 1 - \frac{x}{2} {\rm ln\,} \left|{\frac{1+x}{1-x}}\right| \,,
\eeq
and we note that we can substitute $1 + n_B(q x) \to \frac 12$ due to the symmetries of the integrand in the domain of integration.

{At this point one could worry whether there could be problems of non-integrability in the computation due to
the chiral plasma instability. However, exactly as it occurs in the computation of the energy loss in anisotropic
QED and QCD plasmas \cite{Romatschke:2003vc,Romatschke:2004au}, which exhibit the well-known Weibel instabilties, 
%we find an effective dynamical shielding of the unstable modes. 
we find that the poles associated with the unstable modes are dynamically shielded.
The computation of other quantities such as the momentum broadening would however be more problematic, as there the unstable modes are not shielded
\cite{Romatschke:2006bb,Baier:2008js}, and would be drastically affected by the instability}.

One can easily recognize that the resulting integrals contain both  linear and quadratic ultraviolet divergencies, as opposed to the
$\mu_5 =0$ case, where the leading result in the soft region contains only a logarithmic ultraviolet divergence.
In this case, it turns out to be very convenient to use dimensional regularization. The linear divergence is then set to zero in DR,
and only yields a finite result proportional to $ e^2 m^2_D {e^2\mu_5} / {E}$ (see App.~\ref{app:integralsDR}). The dominant term of $\Delta^{\rm soft}$ is then
provided by the piece that is quadratically divergent: in $d = 3 + 2 \epsilon$ spatial dimensions

\be
\label{softDelta1}
 \Delta^{\rm soft}  \approx  \frac{e^2}{8 \pi v^3 }\frac{e^2\mu_5}{E} \int^{v}_{-v} d x \,x^2\left( 1 - \frac{x^2}{v^2} \right)^{(d-3)/2}  \frac{ v^2-x^2}{1 - x^2} \sum_{h= \pm} 
{\cal I}_q^h (x) \ ,
\ee
with
\be
{\cal I}_q^h (x)= \nu^{3 -d} F(d) \int^\infty_0  {dq} \frac{ q^{d+2}}{\left[ 2 q^2 +  \frac{m_D^2}{1-x^2} -  M^2_h Q_1(x) \right]^2 + \frac{M_h^4 \pi^2 x^2}{4}} \ ,
\ee
where $\nu$ is the {DR} scale, and $F(d)$ is a normalization factor in $d$ dimension, see Eq.~(\ref{nor-fac-d}). 
 A closed analytical expression for the above integral can be obtained if we neglect the
$e^2 \mu_5$ pieces in the denominator,  giving (see App.~\ref{app:integralsDR})
\be
{\cal I}_q^h (x) \approx \frac{m^2_D}{4 \pi^2} A(x) \left(\frac{1}{\epsilon} + \log{\frac{\nu^2}{m^2_D}} + \gamma - \log{4 \pi} \right) \,,
\label{eq:iqhDR}
\ee
with $A(x) = \frac 12 \left( \frac{1}{1-x^2} - Q_1(x) \right)$.
Thus 

\be
 \Delta^{\rm soft}  \approx \frac{e^4 m^2_D}{2 v^3} \frac{\mu_5}{E} \left(\frac{1}{\epsilon} + \log{\frac{\nu^2}{m^2_D}} + \gamma - \log{4 \pi} \right)  f(v) \,,
\ee
where 
\be
f(v) = 2\int^{v}_{-v} d x\,  x^2 \frac{v^2 -x^2}{1 -x^2} A(x) \ 
\ee
is a positive function for $v \in [0,1]$.

The divergence of $ \Delta^{\rm soft}$ in the limit $\epsilon \to 0$ should be cancelled by a hard contribution, as it occurs with the leading
logarithmic divergence (see Appendix \ref{app:DR}). This would require to compute perturbative corrections to the
leading term we computed in the previous section, to find a quadratic infrared divergent piece. Using DR one then expects
to eliminate the pole $1/\epsilon$ and the dependence on the scale $\nu$, resulting in a contribution
$\Delta \propto e^4 \log(e^2) m^2_D \frac{\mu_5}{E} $,
 which would be a correction to the leading result computed in
the previous section. 
{It could also be possible that at this higher order of the computation the perturbative resummed theory breaks down. We defer the investigation of this issue to future projects, as this would be in any case a subleading correction.}

\section{Conclusions}
\label{sec:conclusions}

We have computed the collisional energy loss of an {energetic} massive fermion crossing a chiral plasma with an imbalance of its left-handed and right-handed populations.
In the presence of a chiral imbalance, the energetic fermion interacts differently with the left-handed and right-handed components of the plasma, generating 
new contributions to the energy loss.
These contributions depend on the helicity of the fermion, and we isolated them by focusing on the difference between the energy loss of the two  opposite helicities. {Such contributions can only be due to {\it parity} breaking effects in the medium, and we single them out to study whether they can be used
to quantify the chiral imbalance of the plasma.}

We find that the leading contribution to these helicity-dependent energy loss contributions comes from the exchange of hard photons with the medium constituents, and in some scenarios can become comparable to the leading-order result for a plasma without any chiral imbalance. More specifically, we find that for a very energetic fermion ($v\to 1$) our correction is
 $\Delta \sim e^4 T \mu_5$, compared to the known $ \sim e^4 \log(e^2) T^2$ result for a vanishing $\mu_5$ \cite{Braaten:1991jj}, whereas at smaller velocities we get a  $\mu_5/E$ suppression compared to the leading result.
 
On the other hand, when it comes to softer photon exchanges the chiral imbalance of the medium is felt via a different interaction with the right and left circular polarization components of the photon, which can be incorporated via an extension of the HTL resummation via the inclusion of an anomalous hard dense fermionic loop contribution proportional to the chiral imbalance of the medium  \cite{Laine:2005bt,Akamatsu:2013pjd}. This can be seen to be a subleading effect \cite{Carignano:2018thu}, { perturbatively suppressed as compared with  the contribution
from the hard photon exchange.}

Recall  that chiral plasmas with an imbalance of right and left-handed populations exhibit a chiral plasma instability \cite{Laine:2005bt,Akamatsu:2013pjd},
with a time scale  {$t_{\rm ins} \sim T^2/e^4 \mu^3_5$}, and that our  
 calculation for the fermion energy loss will be valid in a regime before the instability sets in. 
In practice,  if we assume that 
the medium has a length of order $L$, we impose $L/v \ll t_{\rm ins}$.
The time scale should definitely be large for the QED plasma considered in this work.

While we have focused our computations to an electromagnetic plasma, it is
easy to generalize them to QCD { \cite{Braaten:1991we,Peshier:2006hi,Peigne:2008wu,Peigne:2008nd}.}
{First, one} has to take into account 
the proper color and flavor factors in the {corresponding scattering matrix.}
{Note that in QCD there is an additional diagram contributing to the hard sector of the energy loss, namely the collision of the energetic massive quark with the gluons of the medium, but it would not yield an helicity dependent piece in the energy loss, as one assumes that all gluons of different polarizations are equally thermally distributed. 
Then for a heavy quark traversing a chiral QCD plasma, composed by $N_f$ light quark flavors with a chiral chemical potential $\mu_5$, we can take our QED result,
and simply replace the electromagnetic coupling constant by the strong coupling constant $e \rightarrow g$, and take into account a $2 N_f/3$ global factor  },
\be
\Delta^{(QCD)}_{(E \ll M^2/T)}  =  \frac{g^4 T^2 N_f}{18 \pi} \frac{\mu_5}{E} \Big( 1 + \frac{\mu_5^2}{\pi^2 T^2} \Big)
\Big[ \frac{3 v - 5 v^3}{v^3(v^2-1)} - \frac{3}{2v^3}  (v^2 -1) \log\frac{1+v}{1-v} \Big] \,, 
\ee
while for $v=1$  ( $E \gg M^2/T$ ), in the limit  $\mu_5 \ll T$ the leading contribution is given by 
\be
\Delta^{(QCD)}_{{(v = 1)}}  \approx \frac{g^4 N_f}{3\pi^3}\mu_5 \, T \, {\log 2} +  {\cal O}\Big(\frac{\mu_5}{T}, \frac{\mu_5}{E} \Big)  \,.
\ee
{One should as well consider that the chiral plasma instability  in the chromo-electromagnetic fields in}  this case should occur at {$t_{\rm ins} \sim T^2/g^4 \mu^3_5$}.

Our computations should be completed with the evaluation of an helicity dependence of the radiative energy loss. For very heavy fermions, it is known that collisional loss dominates
over radiative loss. It might be particularly interesting to study the case of radiative energy loss for the $v \to 1$ case.
 This last computation  would be needed to answer the question whether by analyzing the helicity dependence 
 the energy loss  of  light  energetic fermions one could get information of the chiral misbalance produced,
for example,  in heavy ion collisions. 

\section*{Acknowledgements} 

We thank J. Soto for instructive conversations on  dimensional regularization,  M. Carrington and 
 D. d'Enterria for general discussions, and J.M. Torres-Rinc\'on for a critical reading of the manuscript. We have been supported by MINECO
(Spain) under the projects FPA2016-81114-P and  PID2019-110165GB-I00 (MCI/AEI/FEDER, UE),  as well as by the project  2017-SGR-929  (Catalonia).  S.C. has also been supported by the MINECO (Spain) under the projects FPA2016-76005-C2-1-P and PID2019-105614GB-C21, and by the 2017-SGR-929 grant (Catalonia). This work was also supported by the COST Action CA15213 THOR.

 \appendix

\section{Collisional Energy loss in a symmetric plasma using dimensional regularization}
\label{app:DR}

Most calculations of the energy loss of a fermion in a plasma (see eg. Refs.\cite{Braaten:1991jj,Peigne:2007sd}) have been performed by regularizing 
the intermediate results with a sharp cutoff separating  the soft and hard pieces of the computation.  The two results are then matched and
the final cutoff dependence disappears.  

The use of a sharp cutoff however has some disadvantages. It does not preserve gauge invariance, and makes the cancellation of power-like divergencies more subtle, should they appear.  It may thus be preferable to use a more refined method. 

In this appendix, we apply dimensional regularization to derive the leading-order result for the energy loss of a massive fermion in a plasma.  

Working in $d = 3 + 2\epsilon$ spatial dimensions, our integrals are modified as 
\be
\int \frac{d^3 q}{(2\pi)^3} \rightarrow \int \frac{d^d q}{(2\pi)^d} =
F(d)    \,\int_0^\infty dq \, q^{d-1} \int_{-1}^1 d\cos\theta \sin^{d-3}\theta \ , 
\ee
with 
\be
\label{nor-fac-d}
F(d)= \frac{ 4 }{(4\pi)^{\frac{d+1}{2}} \Gamma( \frac{d-1}{2})} {= \frac{1}{4\pi^2} + {\cal{O}}(\epsilon) } \,,
\ee
where $\theta$ parametrizes an angle with respect to an external vector, and $\Gamma(z)$ stands for the Gamma function.
Furthermore, in $d$ dimensions one has to change the coupling constant as $e^2 \rightarrow e^2 \nu^{3-d}$, where $\nu$ 
is an auxiliary scale introduced by DR. 

In a similar spirit to previous calculations, we then compute separately the contributions at the soft and hard scales, which will exhibit ultraviolet and infrared divergencies, respectively, appearing
as simple poles $1/\epsilon_{UV}$ and $1/\epsilon_{IR}$. 
The two results are then matched at the scale $\nu$, with the identification 
 $\epsilon_{IR} = \epsilon_{UV}$, and the poles as well as all dependence on the scale $\nu$ cancel out.

{Dimensional regularization has several advantages over cutoff regularization (see \cite{Manohar:2018aog} for a general discussion on DR and effective field theories, or 
\cite{Escobedo:2008sy,Escobedo:2010tu,Escobedo:2011ie} for explicit applications for thermal plasmas). In DR scaleless integrals vanish, and there are
no power divergences. Evaluating integrals using DR is basically the same as evaluating integrals using the method of residues \cite{Manohar:2018aog}, the
result is given in terms of the residues of these poles, which only depend on the physical scales of the theory.

As we will show in our explicit computations of the energy loss in the soft sector,   even if the integrals are
carried out for all momenta, the integral is dominated by $e T$, the soft scale associated with the Debye mass, as it is the only scale in the corresponding integral.
In an analogous way, in the hard sector, the integrals are dominated by contributions at the scale given by the temperature $T$, as can be seen by the presence of the fermionic occupation numbers.

 Let us illustrate how this works by repeating the computation \cite{Braaten:1991jj}  using DR.

\subsection{Soft contribution} 
\label{app-soft}

First we compute the soft contribution to the energy loss using the HTL effective theory.
 Within dimensionally regularized resummed effective field theory the computation will have UV divergences, which appear as simple poles
 in our result.

Generalizing the result of \cite{Braaten:1991jj} to $d$ dimensions, we have
\ba
- \frac{ d E}{d x} \Bigg |_{\rm soft} &= & \frac{ e^2 \nu^{3-d}}{v^2} F(d) \,\int_0^\infty dq \, q^{d-1} \int_{-v}^v dx \,\Big(1 - \frac{x^2}{v^2}\Big)^{(d-3)/2}
\nonumber
\\
& \times & [ 1 + n_B(q x) ] (q x)
\Big(\rho^d_L( q x, q) + (v^2 -x^2) \rho^d_T (q x, q)  \Big)  \,,
\label{eq:dedxRho1}
\ea
where the factor $(1 - \frac{x^2}{v^2})^{(d-3)/2}$ stems from the angular integral in $d$ dimensions, and
 $\rho^d_{L/T}$ are the HTL spectral functions, 
 which can be defined in Coulomb gauge as in the $d=3$ case 
starting from 
the imaginary part of the retarded propagators:
\be
\rho_{L/T}^d (q_0,q) = 2 \, {\rm Im }\, D^d_{L/T} (q_0 + i \epsilon, q) \,.
\ee

 One can substitute $1 + n_B(q x) \to \frac 12$ due to the symmetries of the integrand in the domain of integration in Eq.~(\ref{eq:dedxRho1}). Then the only physical scale
that appears in Eq.~(\ref{eq:dedxRho1}) is the Debye mass.

Expressions of the HTL in $d$ dimensions can be found in \cite{Carignano:2017ovz,Carignano:2019ofj}. 
%\be
%\Pi^{\mu\nu}_d(L) =  2e^2\nu^{3-d} \int\frac{d^{d}k}{(2\pi)^{d}} \frac{ 1- 2 n_F(k) }{k} \left( \frac {  v^\mu v^\nu L^2}{ (v \cdot L)^2} - \frac{v^\mu L^\nu + v^\nu L^\mu}{ v \cdot L}+ g^{\mu \nu} \right) \ ,
%\label{PiHTL}
%\ee
%%
The result is finite and does not present any sort of divergencies, as the scale-less $T=0$ integrals are zero in DR. One thus typically 
computes the HTL diagrams setting $\epsilon=0$ (ie. $d=3$). 
However, since in our calculation they are multiplied by $1/\epsilon$ poles, one should keep pieces up to order $\epsilon$, which can be seen 
as corrections to the Debye mass as well as the Landau damping pieces, as they can give rise to finite pieces in the final result.
In the following, we will mainly focus on the form of the pole obtained within the DR calculation, so we will not derive these explicitly.

Performing the $q$ integral (see Appendix \ref{app:integralsDR} for the explicit results), one finds, in the limit $\epsilon \rightarrow 0$,

\be
-\frac{ d E}{d x} \Bigg |_{\rm soft} =-  \frac{e^2 m^2_{D(3+2\epsilon)}}{16 \pi} \left\{ \left( \frac {1}{v} - \frac{1 -v^2}{2v^2} \ln{\frac{1+v}{1-v}} \right) \left(\frac 1\epsilon - \ln{\frac{{\bar \nu}^2 }{m^2_D}} \right) +
A(v) + A^{\rm extra}(v)  \right\} \,,
\ee 
where ${\bar \nu}^2 = \nu^2 (4 \pi e^{-\gamma})$,  $m^2_{D(3+2\epsilon)}$ denotes the Debye mass in $d=3+2\epsilon$ dimensions keeping pieces up to ${\cal O}(\epsilon)$, which is given by 
(for simplicity we restrict ourselves to $\mu=0$)
\begin{align}
m_{D(3+2\epsilon)}^{2} & = {16e^2}{F(3+2\epsilon)} \nu^{-2\epsilon} \int_0^\infty dk k^{1+2\epsilon} n_F(k) \nn\\
&= m_D^2 \Big[ 1 + \epsilon\Big( 2 - \gamma_E + 2 \frac{\zeta'(2)}{\zeta(2)} - \log\frac{\pi \nu^2}{T^2}  \Big) \Big] = m_D^2 + \epsilon (\delta m_D^2)  \,,
\label{eq:mdDR} 
\end{align}
where $n_F$ is the fermionic occupation number, $\zeta$ denotes the Riemann zeta function, and $\zeta'$ its derivative.

The finite pieces are given by
\ba
\label{finite}
A(v) v^2 &=& \int^v_{-v}dx x^2 \ln{ \Big(1 - \frac{x^2}{v^2} \Big)} \left( 1+ \frac 12 \frac{v^2 -x^2}{1- x^2} \right) \\
\nonumber
&+ & \frac 12 \int^v_{-v}dx x^2 \left( \ln{ (Q^2_1(x) + \frac{\pi^2 x^2}{4} )} + \frac 12 \frac{v^2 -x^2}{1- x^2}  \ln{ (Q^2_2(x) + \frac{\pi^2 x^2}{16} )} \right) \\
\nonumber
&+&  \int^v_{-v}dx x^2 \left( \frac{2Q_1(x)}{\pi x} \arccos{\frac{\pi x}{2 Q_1(x)}} + \frac 12 \frac{v^2 -x^2}{1- x^2}  \frac{Q_2(x)}{\pi x} \arccos{\frac{\pi x}{ Q_2(x)}}  \right) \,,
\ea
{which matches the finite pieces of the soft contribution of the energy loss computed with a cutoff of Ref.~\cite{Braaten:1991jj}, 
and $A^{\rm extra}$ are additional pieces arising from considering HTLs in $d=3+ 2 \epsilon$ dimensions.

Focusing on the pole, we get 
\be
-\frac{ d E}{d x} \Bigg |^{\rm pole}_{\rm soft} =  -\frac{e^2 m^2_D}{16 \pi} \left( \frac {1}{v} - \frac{1 -v^2}{2v^2} \ln{\frac{1+v}{1-v}} \right) \frac 1\epsilon \,, %
  \label{eq:softPole}
  \ee
  and we need to check if it cancels with the hard part.

\subsection{Hard contribution} 
\label{app-hard}

The computation above has to be matched with that of the tree-level scattering in QED,
which we also carry out in $d=3 + 2 \epsilon$ spatial dimensions. This one contains IR divergencies, which are again regulated with DR.

Restricting ourselves to the leading terms in the $1/E$ expansion, 
and considering $\mu=0$ for simplicity, we have \cite{Braaten:1991jj}

\ba
- \frac{ d E}{d x} \Bigg |_{\rm hard} &= & \frac{4 \pi e^4 \nu^{6 -2 d}}{v}  \int \frac{d^d k} 
{(2\pi)^d}  n_F(k)
\int \frac{d^d q}{(2\pi)^d} \int d \omega \, \delta ( \omega + k -|{\bf k}+ \q|) \delta ( \omega - {\bf v} \cdot \q)
\nonumber \\
 & \times &\frac{\omega}{Q^4} \frac{1} {k  |{\bf k}+ \q|} \left( 2 (k - {\bf v} \cdot {\bf k})^2 + \frac {1 -v^2}{2} Q^2 \right)  \,,
\ea
where $n_F$ is the fermionic occupation number.
At this point we can perform an average of the above expression on the directions of ${\bf v}$.

 The angular integral will give us new integration boundaries, as well as an additional factor from the integration
measure in $d$ dimensions. We get to an expression which can be split into two integrals over different boundaries for $q$ and $\omega$:
%$S_d = 2 \pi^{d/2}/\Gamma(d/2) = 4\pi + {\cal{O}}(\epsilon) $
%
\begin{align}
- \frac{ d E}{d x} \Bigg |_{\rm hard} &=  \frac{2 e^4 \nu^{6 -2 d}}{v^2}  
%\frac{\Gamma{(d/2)}}{\sqrt{\pi} \Gamma(\frac{d-1}{2})}  
%\frac{S_d F(d)}
\frac{2\pi^{d/2}}{\sqrt{\pi} \Gamma(\frac{d-1}{2})}   
\frac{F(d)}
{(2\pi)^{d-1}} \int_0^{\infty} d k k^{d-3} n_F(k) \left \{
\int_0^{2k/(1+v)}  dq \, q^{d-5} \int^{v q}_{- vq } d \omega \, \omega \right. \nonumber\\
&+
\left.
\int_{2k/(1+v)}^{2k/(1-v)}  dq \,q^{d-5} \int^{v q}_{q - 2k } d \omega \, \omega \right \}
\left ( \frac {1 -v^2}{2} \frac{q^2}{Q^2} + \frac{3 \omega^2-v^2 q^2}{4q^2} + \frac{3 (\omega k + k^2)}{q^2}
\right.
\nonumber
\\
&-
\left.
 (v^2-1) \frac{\omega k + k^2}{Q^2} \right) \left( 1 - \frac{\omega^2}{v^2 q^2} \right)^{(d-3)/2}  \left[ 1 - \Big( \frac{Q^2}{2 kq} +\frac{\omega}{ q}\Big)^2\right]^{(d-3)/2} \equiv I_1 + I_2 \,.
 \label{eq:dedxdrh}
\end{align}

Since our main focus in this section is to discuss the implementation of dimensional regularization, we focus on the possibly diverging pieces,
which are restricted to the first $q$ integral above, which is IR divergent. The remaining pieces are the same as those obtained in \cite{Braaten:1991jj,Peigne:2007sd}.

Focusing on the first integral of \Eq{eq:dedxdrh} and setting $d=3+2\epsilon$ we have
\begin{align}
I_1 & = \frac{e^4 \nu^{-4 \epsilon}}{v^2\pi^2}  \frac{1}{(4 \pi)^{1+2 \epsilon} }\frac{1}{(\Gamma(1+\epsilon))^2}  \int_0^{\infty} d k k^{2 \epsilon} n_F(k) 
\int_0^{2k/(1+v)}  dq \, q^{-2 +2\epsilon} \int^{v q}_{- vq } d \omega  \nn\\
& \left [ \frac {1 -v^2}{2} \frac{q^2}{Q^2} + \frac{3 \omega^2-v^2 q^2}{4q^2} + \frac{3 (\omega k + k^2)}{q^2}
- (v^2-1) \frac{\omega k + k^2}{Q^2} \right]
 \left( 1 - \frac{\omega^2}{v^2 q^2} \right)^{\epsilon}  \left[ 1 - \Big( \frac{Q^2}{2 kq} +\frac{\omega}{ q}\Big)^2\right]^{\epsilon} \,.
\end{align}
If we now restrict ourselves to extracting the pole and the dependence on the scale $\nu$, we can actually drop several finite pieces and work with 
\be
I_1^{\rm pole}  = \frac{e^4 \nu^{-4 \epsilon}}{v^2\pi^2}  \frac{1}{(4 \pi) }  \int_0^{\infty} d k k^{2 \epsilon} n_F(k) 
\int_0^{2k/(1+v)}  dq \, q^{-2 +2\epsilon} \int^{v q}_{- vq } d \omega  \left( \frac{3 \omega^2 k  }{q^2} -(v^2-1) \frac{\omega^2 k }{Q^2} \right) 
 \,,
\ee
where we exploited the symmetry of the integral to drop all odd terms in $\omega$. 
The $\omega$ integral can now be easily carried out, leading to 
\be
I_1^{\rm pole} = \frac{2e^4 \nu^{-4 \epsilon}}
{\pi} F(3+2\epsilon)
 \left(\frac 1v {+} \frac{(v^2-1)}{2v^2}\ln{\frac{1+v}{1-v}}  \right)
 \int_0^{\infty} d k k^{1+2 \epsilon} n_F(k) 
\int_0^{2k/(1+v)}  dq \, q^{-1 -2\epsilon}  \,.
\ee

 Now we have to evaluate the $q$ integral. 
 Due to the finite upper integration limit, this integral does not vanish in DR and gives
 \be
  \nu^{-2 \epsilon} \int_0^{2k/(1+v)}  dq \, q^{-1 +2\epsilon}  = \frac{1}{2 \epsilon} \left( \frac{2k}{\nu(1+v)}\right)^{2 \epsilon} = \frac{1}{2 \epsilon} + \log{\frac{2 k}{\nu(1 +v)}}  +{\cal O}(\epsilon) \,,
 \ee
 so we are left with 
 \begin{align}
I_1^{\rm pole} & = \frac{2e^4}
{\pi} F(3+2\epsilon)
% was: { 4 \pi^3}
 \left(\frac 1v {+} \frac{(v^2-1)}{2v^2}\ln{\frac{1+v}{1-v}}  \right)   
 \nu^{-2 \epsilon} \int_0^{\infty} d k k^{1+2 \epsilon} n_F(k) \Big[ \frac{1}{2 \epsilon} + \log{\frac{2 k}{\nu(1 +v)}} \Big] \nn\\
& = \frac{e^2}{16\pi v}
 \left(1 {-} \frac{1- v^2}{2v}\ln{\frac{1+v}{1-v}}  \right)  \Big[
 \frac{1}{\epsilon}   m_{D(3+2\epsilon)}^2 
  + \frac{8e^2}
{\pi^2} 
  \int_0^{\infty} d k\, k  n_F(k) \log{\frac{2 k}{\nu(1+v)}} \Big] \,,
\end{align}
where we recognized in the first row the same integral as \Eq{eq:mdDR}, giving the Debye mass in $d=3+2\epsilon$ dimensions.

 Restricting ourselves for simplicity to the case $\mu=0$, we get  
 \be
-\frac{ d E}{d x} \Bigg |_{\rm hard} = \frac{e^4 T^2}{48\pi} \Big[\frac{1}{v} + \frac{v^2-1}{2v^2} \ln{\frac{1+v}{1-v}} \Big] \Big[ \frac{1}{\epsilon}  +\delta m_D^2 - \log \frac{\bar\nu^2}{T^2} + \dots \Big]  \,,
  \label{eq:hardPole}
  \ee
  where $\delta m_D^2$ was defined in \Eq{eq:mdDR} and the dots denote additional finite pieces. 
 We can see that the pole thus cancels out exactly with the one of the soft part (\Eq{eq:softPole}), and the  $\nu$ dependence also drops when adding up the two contributions, which combine to the characteristic $\log(e^2)$ factor.
        
   The generalization to finite chemical potential amounts to replacing $T^2 \to T^2 \Big( 1+ \frac{3\mu^2}{\pi^2 T^2}\Big)$ in the pole, while for the finite pieces one obtains significantly more involved expressions, originating from the evaluation of integrals like $\int_0^\infty dk k \log(k)  [ (n_F(k,\mu) +n_F(k,-\mu)  ]$. Remarkably, these are the same integrals appearing 
   in the calculation of finite pieces at finite chemical potential using a sharp cutoff \cite{Vija:1994is}.

 \section{Useful integrals in dimensional regularization}
 \label{app:integralsDR}
 
 We report in this appendix some integrals we encounter during our calculations, which are regularized in $d=3+2\epsilon$ dimensions using DR. Recall that DR sets power-like divergencies to zero, 
 keeping only results which depend on the physical scales of the system. 
 
  The logarithmic {ultraviolet} divergence encountered in the {soft region of the}  calculation for the energy loss in a thermal plasma without chiral imbalance stems from the integral  
 \begin{align}
 \nu^{- 2 \epsilon} & \int_0^\infty d q \frac{q^{3 + 2 \epsilon}}{(q^2 + m_D^2 a)^2 + m_D^4 b^2} = \frac12 \left(\frac{m_D}{\nu}\right)^{2 \epsilon} \int_0^\infty d z \frac{z^{1 +  \epsilon}}{(z + a)^2 + b^2} \nn\\
& =  - \frac{1}{2} \left( \frac{1}{\epsilon} -  \log{\frac{\nu^2}{m_D^2}} + \frac{1}{2} \log{( a^2 +b^2)} +  \, \frac{a}{b} \arccos{ \frac{a}{\sqrt{a^2 + b^2}}}  \right) + {\cal O}(\epsilon) \,, 
 \end{align}
 and shows up as a $1/\epsilon$ pole, together with a logarithmic dependence on the scale $\nu$, which are finally cancelled by similar contributions in the hard sector,
 which appear there as infrared divergent integrals.
 
When evaluating the new contributions arising in a chiral plasma, we deal with a possibly linearly {ultraviolet} divergent piece. {In $d=3+2\epsilon$ dimensions
the corresponding integral reads}
  \begin{align}
  \nu^{- 2 \epsilon} & \int_0^\infty d q \frac{q^{4 + 2 \epsilon}}{(q^2 + m_D^2 a)^2 + m_D^4 b^2} = \frac{m_D}{2} \left(\frac{m_D}{\nu}\right)^{2 \epsilon} \int_0^\infty d z \frac{z^{3/2 +  \epsilon}}{(z + a)^2 + b^2} \nn\\
  & = - \frac{m_D \pi}{2\sqrt{2}}  \frac{2a+ \sqrt{a^2 + b^2}}{(a+{\sqrt{a^2 + b^2})^{1/2}}} + {\cal O}(\epsilon)  \,,
\end{align}
which however turns out to be finite in DR. 

{ In the chiral plasma there is also a quadratic ultraviolet divergence. In  $d=3+2\epsilon$ dimensions one has to evaluate}
 \begin{align}
  \nu^{- 2 \epsilon} & \int_0^\infty d q \frac{q^{5 + 2 \epsilon}}{(q^2 + m_D^2 a)^2 + m_D^4 b^2} = \frac{m_D^2}{2} \left(\frac{m_D}{\nu}\right)^{2 \epsilon} \int_0^\infty d z \frac{z^{2 +  \epsilon}}{(z + a)^2 + b^2} \nn\\
 &  = a m_D^2 \Big( \frac{1}{\epsilon} - \log{\frac{\nu^2}{m_D^2}} + \frac{1}{2} \big[ \log(a^2 +b^2) + \frac{a^2-b^2}{ab}\arctan\frac{b}{a} \big] \Big) + {\cal O}(\epsilon) \,,
 \end{align}
 which results in the $1/\epsilon$ pole appearing in \Eq{eq:iqhDR}, and a {logarithmic dependence on the scale $\nu$, which should be cancelled with
 a contribution of the hard sector, not evaluated in this manuscript.}


\begin{thebibliography}{99}


%\cite{dEnterria:2009xfs}
\bibitem{dEnterria:2009xfs}
D.~d'Enterria,
%``Jet quenching,''
Landolt-Bornstein \textbf{23}, 471 (2010)
doi:10.1007/978-3-642-01539-7\_16
[arXiv:0902.2011 [nucl-ex]].
%254 citations counted in INSPIRE as of 20 Feb 2021

%\cite{CasalderreySolana:2007zz}
\bibitem{CasalderreySolana:2007zz}
J.~Casalderrey-Solana and C.~A.~Salgado,
%``Introductory lectures on jet quenching in heavy ion collisions,''
Acta Phys. Polon. B \textbf{38}, 3731-3794 (2007)
[arXiv:0712.3443 [hep-ph]].
%196 citations counted in INSPIRE as of 25 Feb 2021

%\cite{Majumder:2010qh}
\bibitem{Majumder:2010qh}
A.~Majumder and M.~Van Leeuwen,
%``The Theory and Phenomenology of Perturbative QCD Based Jet Quenching,''
Prog. Part. Nucl. Phys. \textbf{66}, 41-92 (2011)
doi:10.1016/j.ppnp.2010.09.001
[arXiv:1002.2206 [hep-ph]].
%311 citations counted in INSPIRE as of 20 Feb 2021

%\cite{Qin:2015srf}
\bibitem{Qin:2015srf}
G.~Y.~Qin and X.~N.~Wang,
%``Jet quenching in high-energy heavy-ion collisions,''
Int. J. Mod. Phys. E \textbf{24}, no.11, 1530014 (2015)
doi:10.1142/S0218301315300143
[arXiv:1511.00790 [hep-ph]].
%176 citations counted in INSPIRE as of 01 Feb 2021


%\cite{Kharzeev:2013ffa}
\bibitem{Kharzeev:2013ffa}
D.~E.~Kharzeev,
%``The Chiral Magnetic Effect and Anomaly-Induced Transport,''
Prog. Part. Nucl. Phys. \textbf{75}, 133-151 (2014)
doi:10.1016/j.ppnp.2014.01.002
[arXiv:1312.3348 [hep-ph]].
%278 citations counted in INSPIRE as of 09 Feb 2021

%\cite{Kharzeev:2015znc}
\bibitem{Kharzeev:2015znc}
D.~E.~Kharzeev, J.~Liao, S.~A.~Voloshin and G.~Wang,
%``Chiral magnetic and vortical effects in high-energy nuclear collisions\textemdash{}A status report,''
Prog. Part. Nucl. Phys. \textbf{88}, 1-28 (2016)
doi:10.1016/j.ppnp.2016.01.001
[arXiv:1511.04050 [hep-ph]].
%437 citations counted in INSPIRE as of 13 Feb 2021

%\cite{Huang:2015oca}
\bibitem{Huang:2015oca}
X.~G.~Huang,
%``Electromagnetic fields and anomalous transports in heavy-ion collisions --- A pedagogical review,''
Rept. Prog. Phys. \textbf{79}, no.7, 076302 (2016)
doi:10.1088/0034-4885/79/7/076302
[arXiv:1509.04073 [nucl-th]].
%170 citations counted in INSPIRE as of 22 Jan 2021




%\cite{Carignano:2018thu}
\bibitem{Carignano:2018thu}
S.~Carignano and C.~Manuel,
%``Damping rate of a fermion in ultradegenerate chiral matter,''
Phys. Rev. D \textbf{99}, no.9, 096022 (2019)
doi:10.1103/PhysRevD.99.096022
[arXiv:1811.06394 [hep-ph]].
%1 citations counted in INSPIRE as of 22 Jan 2021


%\cite{Romatschke:2003vc}
\bibitem{Romatschke:2003vc}
P.~Romatschke and M.~Strickland,
%``Energy loss of a heavy fermion in an anisotropic QED plasma,''
Phys. Rev. D \textbf{69}, 065005 (2004)
doi:10.1103/PhysRevD.69.065005
[arXiv:hep-ph/0309093 [hep-ph]].
%55 citations counted in INSPIRE as of 21 Apr 2021


%\cite{Romatschke:2004au}
\bibitem{Romatschke:2004au}
P.~Romatschke and M.~Strickland,
%``Collisional energy loss of a heavy quark in an anisotropic quark-gluon plasma,''
Phys. Rev. D \textbf{71}, 125008 (2005)
doi:10.1103/PhysRevD.71.125008
[arXiv:hep-ph/0408275 [hep-ph]].
%67 citations counted in INSPIRE as of 21 Apr 2021


%\cite{Romatschke:2006bb}
\bibitem{Romatschke:2006bb}
P.~Romatschke,
%``Momentum broadening in an anisotropic plasma,''
Phys. Rev. C \textbf{75}, 014901 (2007)
doi:10.1103/PhysRevC.75.014901
[arXiv:hep-ph/0607327 [hep-ph]].
%112 citations counted in INSPIRE as of 21 Apr 2021

%\cite{Baier:2008js}
\bibitem{Baier:2008js}
R.~Baier and Y.~Mehtar-Tani,
%``Jet quenching and broadening: The Transport coefficient q-hat in an anisotropic plasma,''
Phys. Rev. C \textbf{78}, 064906 (2008)
doi:10.1103/PhysRevC.78.064906
[arXiv:0806.0954 [hep-ph]].
%58 citations counted in INSPIRE as of 21 Apr 2021





%\cite{Braaten:1991jj}
\bibitem{Braaten:1991jj}
E.~Braaten and M.~H.~Thoma,
%``Energy loss of a heavy fermion in a hot plasma,''
Phys. Rev. D \textbf{44}, 1298-1310 (1991)
doi:10.1103/PhysRevD.44.1298
%284 citations counted in INSPIRE as of 17 Feb 2021

%\cite{Peigne:2007sd}
\bibitem{Peigne:2007sd}
S.~Peigne and A.~Peshier,
%``Collisional Energy Loss of a Fast Muon in a Hot QED Plasma,''
Phys. Rev. D \textbf{77}, 014015 (2008)
doi:10.1103/PhysRevD.77.014015
[arXiv:0710.1266 [hep-ph]].
%41 citations counted in INSPIRE as of 22 Jan 2021

%%\cite{Romatschke:2003vc}
%\bibitem{Romatschke:2003vc}
%P.~Romatschke and M.~Strickland,
%%``Energy loss of a heavy fermion in an anisotropic QED plasma,''
%Phys. Rev. D \textbf{69}, 065005 (2004)
%doi:10.1103/PhysRevD.69.065005
%[arXiv:hep-ph/0309093 [hep-ph]].
%%55 citations counted in INSPIRE as of 25 Feb 2021

%\cite{Hansen:2020irw}
\bibitem{Hansen:2020irw}
J.~Hansen and K.~Tuchin,
%``Collisional energy loss and the Chiral Magnetic Effect,''
[arXiv:2012.06089 [hep-ph]].
%0 citations counted in INSPIRE as of 23 Feb 2021

%\cite{Tuchin:2018sqe}
\bibitem{Tuchin:2018sqe}
K.~Tuchin,
%``Radiative instability of quantum electrodynamics in chiral matter,''
Phys. Lett. B \textbf{786}, 249-254 (2018)
doi:10.1016/j.physletb.2018.09.055
[arXiv:1806.07340 [hep-ph]].
%8 citations counted in INSPIRE as of 23 Feb 2021


%\cite{Weldon:1983jn}
\bibitem{Weldon:1983jn}
H.~A.~Weldon,
%``Simple Rules for Discontinuities in Finite Temperature Field Theory,''
Phys. Rev. D \textbf{28}, 2007 (1983)
doi:10.1103/PhysRevD.28.2007
%515 citations counted in INSPIRE as of 19 Feb 2021

%\cite{Manuel:2000mk}
\bibitem{Manuel:2000mk}
C.~Manuel,
%``Dispersion relations in ultradegenerate relativistic plasmas,''
Phys. Rev. D \textbf{62}, 076009 (2000)
doi:10.1103/PhysRevD.62.076009
[arXiv:hep-ph/0005040 [hep-ph]].
%54 citations counted in INSPIRE as of 22 Jan 2021

%\cite{Nieves:1988qz}
\bibitem{Nieves:1988qz}
J.~F.~Nieves and P.~B.~Pal,
%``$P$ and \{CP\} Odd Terms in the Photon Selfenergy Within a Medium,''
Phys. Rev. D \textbf{39}, 652 (1989)
[erratum: Phys. Rev. D \textbf{40}, 2148 (1989)]
doi:10.1103/PhysRevD.39.652
%49 citations counted in INSPIRE as of 22 Jan 2021

\bibitem{LeBellac}
M. Le Bellac, ``{\it Thermal Field Theory}'', Cambridge University Press,  Cambridge 1996.

%\cite{Laine:2005bt}
\bibitem{Laine:2005bt}
M.~Laine,
%``Real-time Chern-Simons term for hypermagnetic fields,''
JHEP \textbf{10}, 056 (2005)
doi:10.1088/1126-6708/2005/10/056
[arXiv:hep-ph/0508195 [hep-ph]].
%31 citations counted in INSPIRE as of 22 Jan 2021

%\cite{Akamatsu:2013pjd}
\bibitem{Akamatsu:2013pjd}
Y.~Akamatsu and N.~Yamamoto,
%``Chiral Plasma Instabilities,''
Phys. Rev. Lett. \textbf{111}, 052002 (2013)
doi:10.1103/PhysRevLett.111.052002
[arXiv:1302.2125 [nucl-th]].
%139 citations counted in INSPIRE as of 22 Jan 2021

%\cite{Manuel:2013zaa}
\bibitem{Manuel:2013zaa}
C.~Manuel and J.~M.~Torres-Rincon,
%``Kinetic theory of chiral relativistic plasmas and energy density of their gauge collective excitations,''
Phys. Rev. D \textbf{89}, no.9, 096002 (2014)
doi:10.1103/PhysRevD.89.096002
[arXiv:1312.1158 [hep-ph]].
%59 citations counted in INSPIRE as of 22 Jan 2021

%\cite{Carignano:2019zsh}
\bibitem{Carignano:2019zsh}
S.~Carignano, C.~Manuel and J.~M.~Torres-Rincon,
%``Chiral kinetic theory from the on-shell effective field theory: Derivation of collision terms,''
Phys. Rev. D \textbf{102}, no.1, 016003 (2020)
doi:10.1103/PhysRevD.102.016003
[arXiv:1908.00561 [hep-ph]].
%13 citations counted in INSPIRE as of 22 Jan 2021

%\cite{Carignano:2018gqt}
\bibitem{Carignano:2018gqt}
S.~Carignano, C.~Manuel and J.~M.~Torres-Rincon,
%``Consistent relativistic chiral kinetic theory: A derivation from on-shell effective field theory,''
Phys. Rev. D \textbf{98}, no.7, 076005 (2018)
doi:10.1103/PhysRevD.98.076005
[arXiv:1806.01684 [hep-ph]].
%27 citations counted in INSPIRE as of 03 Mar 2021

%\cite{Braaten:1991we}
\bibitem{Braaten:1991we}
E.~Braaten and M.~H.~Thoma,
%``Energy loss of a heavy quark in the quark - gluon plasma,''
Phys. Rev. D \textbf{44}, no.9, 2625 (1991)
doi:10.1103/PhysRevD.44.R2625
%339 citations counted in INSPIRE as of 23 Feb 2021

%\cite{Peshier:2006hi}
\bibitem{Peshier:2006hi}
A.~Peshier,
%``The QCD collisional energy loss revised,''
Phys. Rev. Lett. \textbf{97}, 212301 (2006)
doi:10.1103/PhysRevLett.97.212301
[arXiv:hep-ph/0605294 [hep-ph]].
%66 citations counted in INSPIRE as of 23 Feb 2021

%\cite{Peigne:2008nd}
\bibitem{Peigne:2008nd}
S.~Peigne and A.~Peshier,
%``Collisional energy loss of a fast heavy quark in a quark-gluon plasma,''
Phys. Rev. D \textbf{77}, 114017 (2008)
doi:10.1103/PhysRevD.77.114017
[arXiv:0802.4364 [hep-ph]].
%126 citations counted in INSPIRE as of 23 Feb 2021

%\cite{Peigne:2008wu}
\bibitem{Peigne:2008wu}
S.~Peigne and A.~V.~Smilga,
%``Energy losses in a hot plasma revisited,''
Phys. Usp. \textbf{52}, 659-685 (2009)
doi:10.3367/UFNe.0179.200907a.0697
[arXiv:0810.5702 [hep-ph]].
%78 citations counted in INSPIRE as of 23 Feb 2021




%\cite{Manohar:2018aog}
\bibitem{Manohar:2018aog}
A.~V.~Manohar,
%``Introduction to Effective Field Theories,''
Les Houches Lect. Notes \textbf{108} (2020)
doi:10.1093/oso/9780198855743.003.0002
[arXiv:1804.05863 [hep-ph]].
%62 citations counted in INSPIRE as of 19 Feb 2021

%\cite{Escobedo:2008sy}
\bibitem{Escobedo:2008sy}
M.~A.~Escobedo and J.~Soto,
%``Non-relativistic bound states at finite temperature (I): The Hydrogen atom,''
Phys. Rev. A \textbf{78}, 032520 (2008)
doi:10.1103/PhysRevA.78.032520
[arXiv:0804.0691 [hep-ph]].
%82 citations counted in INSPIRE as of 22 Jan 2021

%\cite{Escobedo:2010tu}
\bibitem{Escobedo:2010tu}
M.~A.~Escobedo and J.~Soto,
%``Non-relativistic bound states at finite temperature (II): the muonic hydrogen,''
Phys. Rev. A \textbf{82}, 042506 (2010)
doi:10.1103/PhysRevA.82.042506
[arXiv:1008.0254 [hep-ph]].
%35 citations counted in INSPIRE as of 22 Jan 2021

%\cite{Escobedo:2011ie}
\bibitem{Escobedo:2011ie}
M.~A.~Escobedo, J.~Soto and M.~Mannarelli,
%``Non-relativistic bound states in a moving thermal bath,''
Phys. Rev. D \textbf{84}, 016008 (2011)
doi:10.1103/PhysRevD.84.016008
[arXiv:1105.1249 [hep-ph]].
%45 citations counted in INSPIRE as of 22 Jan 2021

%\cite{Carignano:2017ovz}
\bibitem{Carignano:2017ovz}
S.~Carignano, C.~Manuel and J.~Soto,
%``Power corrections to the HTL effective Lagrangian of QED,''
Phys. Lett. B \textbf{780}, 308-312 (2018)
doi:10.1016/j.physletb.2018.03.012
[arXiv:1712.07949 [hep-ph]].
%5 citations counted in INSPIRE as of 22 Jan 2021

%\cite{Carignano:2019ofj}
\bibitem{Carignano:2019ofj}
S.~Carignano, M.~E.~Carrington and J.~Soto,
%``The HTL Lagrangian at NLO: the photon case,''
Phys. Lett. B \textbf{801}, 135193 (2020)
doi:10.1016/j.physletb.2019.135193
[arXiv:1909.10545 [hep-ph]].
%4 citations counted in INSPIRE as of 22 Jan 2021

%\cite{Vija:1994is}
\bibitem{Vija:1994is}
H.~Vija and M.~H.~Thoma,
%``Braaten-Pisarski method at finite chemical potential,''
Phys. Lett. B \textbf{342}, 212-218 (1995)
doi:10.1016/0370-2693(94)01378-P
[arXiv:hep-ph/9409246 [hep-ph]].
%81 citations counted in INSPIRE as of 22 Jan 2021




\end{thebibliography}
\end{document}